\documentclass[10pt,twocolumn,letterpaper]{article}

\usepackage{wacv}              
\usepackage[accsupp]{axessibility}

\usepackage[utf8]{inputenc}   
\usepackage[T1]{fontenc}      

\usepackage[table]{xcolor}
\definecolor{radHeaderBlue}{HTML}{1F4E79}
\definecolor{radHeaderLight}{HTML}{D9E2F3}
\definecolor{radStripe}{HTML}{F5F7FB}

\usepackage{graphicx}         
\usepackage{subcaption}       

\usepackage{booktabs}         
\usepackage{multirow}         
\usepackage{makecell}         
\usepackage{tabularx}         
\usepackage[flushleft]{threeparttable} 
\definecolor{lightgray}{gray}{0.9}     

\usepackage{siunitx}          

\usepackage{algorithm}
\usepackage{algorithmic}

\usepackage[none]{hyphenat}   
\usepackage{textcomp}         
\usepackage{url}              
\usepackage{verbatim}         
\usepackage{pifont}           
\usepackage{stfloats}         

\usepackage[most]{tcolorbox}

\usepackage{tikz}
\usetikzlibrary{positioning,shapes,arrows}

\usepackage{array}

\definecolor{wacvblue}{rgb}{0.21,0.49,0.74}
\usepackage[pagebackref,breaklinks,colorlinks,allcolors=wacvblue]{hyperref}

\def\wacvPaperID{1685} 
\def\confName{WACV}
\def\confYear{2026}

\title{SSMRadNet : A Sample-wise State-Space Framework for Efficient and Ultra-Light Radar Segmentation and Object Detection}

\author{
Anuvab Sen, Mir Sayeed Mohammad, Saibal Mukhopadhyay\\
Georgia Institute of Technology, Atlanta, Georgia, USA\\
{\tt\small asen74@gatech.edu, mirsayeedmohammad@gatech.edu, saibal.mukhopadhyay@ece.gatech.edu}
}

\begin{document}
\maketitle

\begin{abstract}
We introduce \textit{SSMRadNet}, the first multi-scale State Space Model (SSM) based detector for Frequency Modulated Continuous Wave (FMCW) radar that sequentially processes raw ADC
samples through two SSMs. One SSM learns a chirp-wise feature by sequentially processing samples from all receiver channels within one chirp, and a second SSM learns a representation of a frame by sequentially processing chirp-wise features. The latent representations of a radar frame are decoded to perform segmentation and detection tasks. 
Comprehensive evaluations on the RADIal dataset show SSMRadNet has \textbf{10-33× fewer parameters} and \textbf{60-88× less computation (GFLOPs)} while being \textbf{3.7× faster} than state-of-the-art transformer and convolution-based radar detectors at competitive performance for segmentation tasks. 
\end{abstract}   
\section{Introduction} \label{sec:intro}

Frequency Modulated Continuous Wave (FMCW) radars improve the perception of autonomous systems under adverse conditions \cite{patole2017automotive,yao2023radarcamreview}.
A FMCW radar frame is represented by a $C \!\times\! S \!\times\! N_{R_x}$ tensor of digitized samples (analog-to-digital-converters (ADC) cube), where $C$, $N_{R_x}$, and $S$ are the number of chirps per frame, ADCs (receiver channels), and samples per chirp, respectively. Higher $N_{R_x}$, $C$, and sampling bandwidth (more $S$ per chirp) improve angular, velocity, and range resolutions, respectively, but also increase computational demand on perception algorithms.

\begin{figure*}
    \centering
    \includegraphics[width = \linewidth]{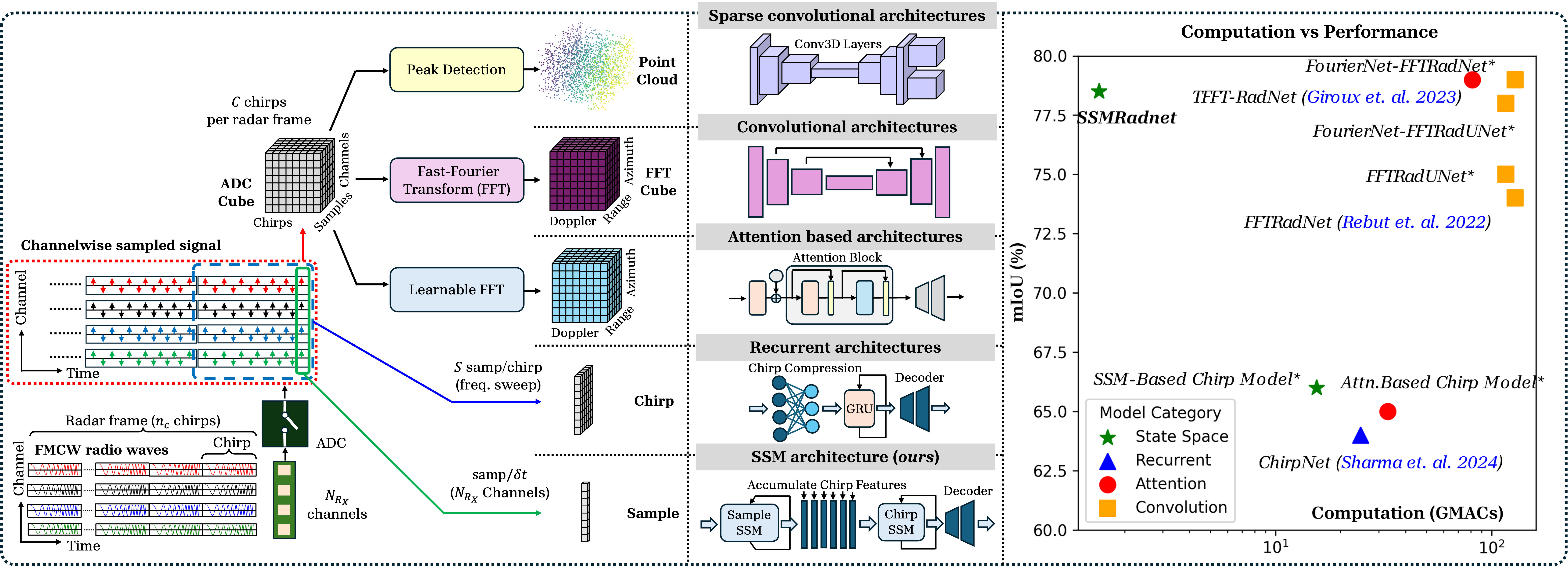}    \caption{\textbf{Motivation from Prior Works:} Traditional convolutional\cite{rebut2022radial}, attention-based\cite{giroux2023tfftradnet} and recurrent\cite{sharma2024chirpnet} networks are computationally expensive because they rely on a large volume of data (\textit{left}). Our state space modeling approach scales linearly in computation with increased radar resolution ($C\times S\times N_{R_X}$) because of our sample by sample processing, achieving state of the art performance for a fraction of the computation (\textit{right}).  (*hybrid models; refer \Cref{tab:results}).}
    \label{fig:motivation}
\end{figure*}

Traditionally, radar perception models generate sparse point clouds from a 3D Range-Azimuth-Doppler (RAD) tensor, which are computed by applying multi-stage FFTs on ADC cubes. The point clouds are processed by Convolutional Neural Networks (CNN) for end tasks (Figure~\ref{fig:motivation},\Cref{tab:comparison}) \cite{richards2014fundamentals,skolnik2008radar,patole2017automotive, schmidt1986music,roy1989esprit, major2019vehicle,skolnik2008radar, scheiner2021comparison,yang2020radarnet}. However, creation and processing of point clouds introduce computation and latency bottlenecks. 

To address the point cloud challenges, recent works have demonstrated convolution, attention, and recurrent network based approaches to learn features directly from ADC cube (\Cref{tab:comparison}) \cite{yang2020radarnet,wang2021rodnet,meyer2019deepradarcam,wu2024diffusion,chu2023mtdetr,zhao2023cubelearn,giroux2023tfftradnet,liu2021swin}.  
As an example, Rebut et al.\ pioneered FFT-RadNet\cite{rebut2022radial} that avoids computing the full 3D range-azimuth-Doppler tensor by using a CNN that learns to recover object angles from the 2D range-Doppler input, thus eliminating one FFT stage and reducing computation. 
As an alternative, ADCNet \cite{zhang2023adcnet} learns representations directly from the ADC cube using CNN based approaches further reducing computation. However, parameter, computation, and memory complexity of 3D CNN increases non-linearly with dimension of the ADC cube.
T‑FFT‑RadNet family \cite{giroux2023tfftradnet,liu2021swin} proposed transformer based methods that interpret samples within the ADC cube as tokens and replace CNNs with attention. The design shows improved performance (accuracy), but as the computational cost of self‑attention scales quadratically, processing long radar sample‑by‑chirp sequences is computationally expensive (\Cref{tab:comparison}).


ChirpNet pioneered a sequential method where digitized ADC samples from all receivers for a chirp are encoded into a feature vector, and features across chirps are learned using a recurrent network \cite{sharma2024chirpnet}. The design shows lower computation, parameters, and latency than prior methods, but with a lower perception accuracy (\Cref{tab:comparison}).


\begin{table*}[t]
\footnotesize
\centering
\caption{Radar object detection architectures; Trade-offs:
\textcolor{green}{\checkmark} (desirable),
\textcolor{red}{\ding{55}} (undesirable),
\textcolor{orange}{\textbullet} (moderate).}
\label{tab:comparison}
\setlength{\tabcolsep}{3pt}%
\begin{tabularx}{\textwidth}{
  @{}
  >{\raggedright\arraybackslash}p{1.8cm}
  >{\raggedright\arraybackslash}p{2.8cm}
  >{\raggedright\arraybackslash}p{2.0cm}
  >{\raggedright\arraybackslash}p{1.8cm}
  >{\raggedright\arraybackslash}p{3.2cm}
  >{\raggedright\arraybackslash}p{1.8cm}
  >{\raggedright\arraybackslash}p{1.6cm}
  >{\raggedright\arraybackslash}p{1.1cm}
  @{}
}
\toprule
\textbf{Class} & \textbf{Model} & \textbf{Input Type} & \textbf{Feature Shape} &
\textbf{Processing Core} & \textbf{Params (M)} & \textbf{Computation (GMACs)} & \textbf{Latency} \\
\midrule
\multirow{3}{*}{\textbf{Convolution}} &
Point-cloud Conv Detectors \cite{yang2018pixor} &
Point Cloud (CFAR) & Sparse 3D tensor &
CNN / FPN over sparse voxel / BEV grids &
High \textcolor{red}{\ding{55}} &
High \textcolor{red}{\ding{55}} &
High \textcolor{red}{\ding{55}} \\
\cmidrule(r){2-8}
& FFT-RadNet \cite{rebut2022radial} &
Range--Doppler (RD) & 2D tensor &
CNN encoder + FPN angle regression &
High \textcolor{red}{\ding{55}} &
High \textcolor{red}{\ding{55}} &
High \textcolor{red}{\ding{55}} \\
\cmidrule(r){2-8}
& ADCNet (Conv) \cite{zhang2023adcnet} &
Raw ADC $\rightarrow$ RD & 2D tensor &
Learnable signal proc + Conv blocks &
High \textcolor{red}{\ding{55}} &
High \textcolor{red}{\ding{55}} &
High \textcolor{red}{\ding{55}} \\
\midrule
\multirow{2}{*}{\textbf{Attention}} &
T-FFT-RadNet (Swin) \cite{giroux2023tfftradnet,liu2021swin} &
RD / RAD / raw ADC & 2D / 3D tensor &
Hierarchical Swin self attention + heads &
High \textcolor{red}{\ding{55}} &
High \textcolor{red}{\ding{55}} (quadratic) &
High \textcolor{red}{\ding{55}} \\
\cmidrule(r){2-8}
& RadarFormer \cite{Dalbah2023RadarFormer} &
RA / RD (merged) & 2D tensor &
Channel--chirp--time merging transformer &
High \textcolor{red}{\ding{55}} &
High \textcolor{red}{\ding{55}} &
High \textcolor{red}{\ding{55}} \\
\midrule
\multirow{1}{*}{\textbf{Recurrent}} &
ChirpNet \cite{sharma2024chirpnet} &
Raw ADC chirp / ($R_x$) & 2D tensor &
Sequential GRU + lightweight MLP &
Mid \textcolor{orange}{\textbullet} &
Mid \textcolor{orange}{\textbullet} &
Mid \textcolor{orange}{\textbullet} \\
\midrule
\multirow{1}{*}{\textbf{SSM}} &
SSMRadNet (ours) &
Raw ADC sample / chirp & 1D tensor &
Sample- \& Chirp-wise SSM's &
Low \textcolor{green}{\checkmark} &
Low \textcolor{green}{\checkmark} &
Low \textcolor{green}{\checkmark} \\
\bottomrule
\end{tabularx}
\end{table*} 

To further reduce computation while maintaining accuracy, we propose \textit{SSMRadNet}, a novel radar processing framework that models (synchronous) samples from all receivers ($N_{Rx}$ antennas) as sequential tokens and processes using a multi-scale state-space model (SSM)(Figure~\ref{fig:motivation}). 
The SSM can capture long‑duration dependencies, and its computation scales linearly with sequence length, instead of quadratically, as in transformers \cite{gu2021s4, goel2022s4nd, fang2023selectivessm, chen2024mamba, zhu2024vim}. Although, SSM shows superior performance than transformers on vision tasks\cite{gu2021s4,chen2024mamba}, SSM for radar processing have not been demonstrated yet.


SSMRadNet has one SSM (i.e., states) to process ADC samples within a chirp (the fast‑time axis). The learned representations from successive chirps are processed sequentially using a second SSM (the slow‑time axis) to create a latent map for a frame. The final latent map is decoded for downstream detection or segmentation.  

We demonstrate SSMRadNet on two radar vision datasets, namely, \textsc{RADIal} \& \textsc{RADICal}. As illustrated in Figure~\ref{fig:motivation}, for the segmentation task on RADIal, \textit{SSMRadNet} demonstrates more than $60\times$ reduction in computation but at a competitive performance. The computational advantage originates from eliminating FFT processing, 3D convolution, and quadratic self‑attention bottleneck in transformers. Compared to ChirpNet, \textit{SSMRadNet}, shows more than $15\times$ computation reduction in computation with a substantially better accuracy, thanks to sequentially learning the temporal features between successive ADC samples within a chirp.    


The key contributions of this paper are as follows : 
\\ \textbf{(1) SSM for Radar Processing :} Our method is the first multi-scale SSM framework for radar processing framework to learn features from sequentially obtained ADC samples modeled as tokens. 
\\ \textbf{(2) Sample-by-sample radar detection:} We propose the first sample-by-sample radar data processing architecture for segmentation and detection eliminating the memory and latency of buffering ADC cubes.
\\ \textbf{(3) Compute efficiency at competitive performance:} Our model has significantly lower parameters, computation, and processing latency compared to prior works, but maintains comparable segmentation and detection performance.


\section{Proposed Model Architecture} \label{sec:arch}

\begin{figure*}[htb]
    \centering
    \includegraphics[width = \linewidth]{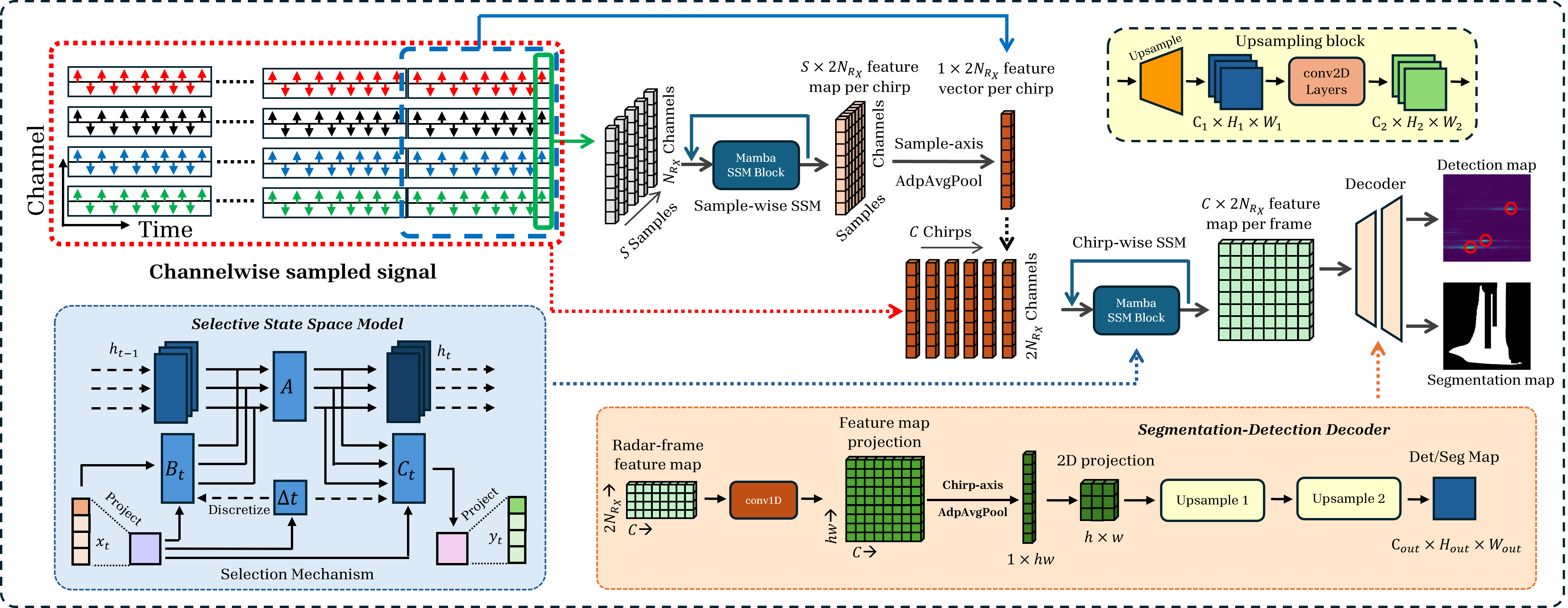}
    \caption{\textbf{SSMRadNet Architecture:} Raw complex ADC samples from $N_{R_X}$-channel feed into sample-SSM block. Sample states/chirp are aggregated into chirp feature maps and temporally pooled into token embeddings. Each chirp token then passes through chirp-SSM blocks to capture inter-chirp dynamics (motion, velocity). Their outputs are aggregated into radar-frame features and decoded—via a channel projection and parallel upsampling+convolutional layers—into detection and segmentation heads.}
    \label{fig:SSMRadNet_arch}

    \label{fig:model_architecture}
\end{figure*}

SSMRadNet is an end-to-end neural architecture that takes raw radar ADC data as input and produces a bird’s-eye-view occupancy map as output. \Cref{fig:model_architecture} illustrates the overall design. The key idea is to model the radar data in sequential form, preserving the temporal and channel information. We decompose the problem into three stages: (1) Per-sample embedding, converting each fast-time ADC sample (across antennas) into a  token representation; (2) SSM sequence modeling, which applies state-space model (Mamba) blocks first along the sample dimension of each chirp (intra-chirp modeling) and then along the chirp dimension (inter-chirp modeling) to capture range and azimuth structure; and (3) spatial decoding, which projects and transforms the sequence outputs into a 2D spatial feature map for the free driving space segmentation and vehicle detection task.

\subsection{Radar Input and Tokenisation}

Raw ADC samples arrive continuously at a fixed “fast‑time” rate.  At each tick
\[
  t = 1,2,\dots,C\times S,
\]
where
- \(C\) = number of chirps per frame,  
- \(S\) = number of fast‑time samples (range bins) per chirp,  

The ADC hardware simultaneously digitizes all \(N_{\mathrm{Rx}}\) receiver channels, producing
\[
  \mathbf{X}(t)
  = \bigl[X_{c,1}(k),\,X_{c,2}(k),\,\dots,\,X_{c,N_{\mathrm{Rx}}}(k)\bigr]^\top
  \in \mathbb{C}^{N_{\mathrm{Rx}}},
\]
\[
\begin{aligned}
  c &= \bigl\lceil t / S \bigr\rceil
       &&\text{(current chirp index)},\\
  s &= t - (c-1)\,S
       &&\text{(fast‑time sample index within chirp \(i\)).}
\end{aligned}
\]
Immediately upon acquisition, we embed each vector:
\begin{equation}\label{eq:embed_time_final}
  \mathbf{z}(t)
  = \mathrm{MLP}\bigl([\Re(\mathbf{X}(t));\,\Im(\mathbf{X}(t))]\bigr)
  \;\in\;\mathbb{R}^{N_{R_X}}
\end{equation}
so that no buffering beyond the current tick is required. These embeddings are passed to the intra-chirp SSM block sequentially.



\subsection{Sample‑wise Mamba Block (Intra‑Chirp SSM)}\label{sec:intra_chirp_ssm}

We leverage the temporal correlations between individual ADC
samples to extract relevant features.  To capture fine‑grained
range‑domain structure within each chirp, we employ a detailed
\textsc{MambaSSM} block that augments the raw token sequence with causal
convolution.
These tokens \(\mathbf{z}_{c,s}\) are split into \(N_{R_X}\) channels
and fed through a grouped causal convolution of width
\(d_{\text{conv}}\):
\begin{equation}\label{eq:causal_conv}
  x_{\mathrm{conv},s}^{(g)}
  \;=\;
  \sum_{k=0}^{d_{\text{conv}}-1}
    w_{g}[k]\,\mathbf{z}_{c,s-k}^{(g)}
  \;+\; b_{g},
  \quad
  s=1,\dots,S
\end{equation}
with zero‑padding for \(s<d_{\text{conv}}\), weights \(w_{g}\), bias
\(b_{g}\). 
The grouped causal convolution maintains a small (first in first out) FIFO buffer of the previous \(d_{\mathrm{conv}}-1\) token embeddings, computing each \(x_{\mathrm{conv},s}\) in realtime.

Next, we linearly project each convolved vector
\(x_{\mathrm{conv},s}\in\mathbb{R}^{N_{R_X}}\) into three modulation streams:
\begin{equation}\label{eq:proj_p}
\mathbf{p}_{s} = W_{p}\,x_{\mathrm{conv},s} = \bigl[D_{\mathrm{dt},s},\,B_{\mathrm{mod},s},\,C_{\mathrm{mod},s}\bigr]
\end{equation}
$D_{\mathrm{dt},s},\,B_{\mathrm{mod},s},\,C_{\mathrm{mod},s}  \in\mathbb{R}^{d_{\mathrm{state}}}$, where
$d_{state}$ determines how many hidden units the SSM uses to accumulate information over the fast-time samples.

A learnable log‐decay parameter \(A_{\log}\in\mathbb{R}^{d_{\text{state}}}\) yields the per‐step decay (“\(\odot\)” is element‐wise multiplication):
\begin{equation}\label{eq:decay}
  A = -\exp\!\bigl(A_{\log}\bigr),
  \quad
  \mathrm{decay}_{t}
  = \exp\!\bigl(dt_{t}\odot A\bigr),
\end{equation}

Here, \(A\in\mathbb{R}^{d_{\mathrm{state}}}\) is the per‑channel decay coefficient that governs how much information of the previous hidden state carries over at each step. 

The internal SSM state \(\mathbf{h}_{s}\in\mathbb{R}^{d_{\text{state}}}\) updates
recursively:
\begin{equation}\label{eq:state_update}
  \mathbf{h}_{s}
  \;=\;
  \mathbf{h}_{s-1}\odot\mathrm{decay}_{s}
  \;+\;
  \tilde{x}_{s}\odot\bigl(dt_{s}\odot B_{\mathrm{mod},s}\bigr),
\end{equation}
with \(\tilde{x}_{s}\) the appropriately reshaped convolved output.
Finally, the block output \(y_{s}\in\mathbb{R}^{d_{\text{state}}}\)
aggregates the state via:
\begin{equation}\label{eq:output}
  y_{s}
  \;=\;
  \sum_{j=1}^{d_{\text{state}}}
      \mathbf{h}_{s}^{(j)}\,C_{\mathrm{mod},s}^{(j)}
  \;+\;
  D\,\tilde{x}_{s},
\end{equation}
where \(D\in\mathbb{R}^{d_{\text{state}}\times N_{R_x}}\) is a learned
skip‑connection matrix.


\medskip
\noindent\textit{\textbf{State summarisation:}}  
We apply average pooling over the hidden states 
\(\mathbf{h}_{0-S}\), effectively
compressing the temporal sequence \((S\times N_{R_X})\) into a
single vector $ y_c \in \mathbb{R}^{N_{R_X}}$.  This compact representation carrying the distilled intra‑chirp information is again passed through an MLP for channel expansion to $ y_{ce} \in \mathbb{R}^{2N_{R_X}}$ and passed to the Chirp-SSM block.

\subsection{Chirp-wise Mamba Block (Inter-Chirp SSM)}\label{sec:inter_ssm}

We apply a similar SSM block from \Cref{sec:intra_chirp_ssm} to the sequence of chirp tokens $\mathbf{y}_{ce} \in \mathbb{R}^{2N_{R_X}}$ along the chirp axis:


\[
\mathbf{u}_c = \text{MambaSSM}(\mathbf{y}_{ce}), \quad c = 1,\dots,C
\]

We take the output $u_c$ for each chirp as they arrive sequentially. These outputs are accumulated over the radar frame to create the slow-time vs channel feature map $U \in \mathbb{R}^{C\times 2N_{R_X}}$ .
\[
  \mathbf{U} \;=\; [\mathbf{u}_{1}, \mathbf{u}_{2}, \dots, \mathbf{u}_{C}]\;\in\;\mathbb{R}^{C\times 2N_{R_X}}
\]

This accumulated temporal feature map $\mathbf{U}$ captures temporal correlations across the chirps.

\begin{figure*}[htb]
    \centering
    \includegraphics[width = \linewidth, height=0.19\textheight]{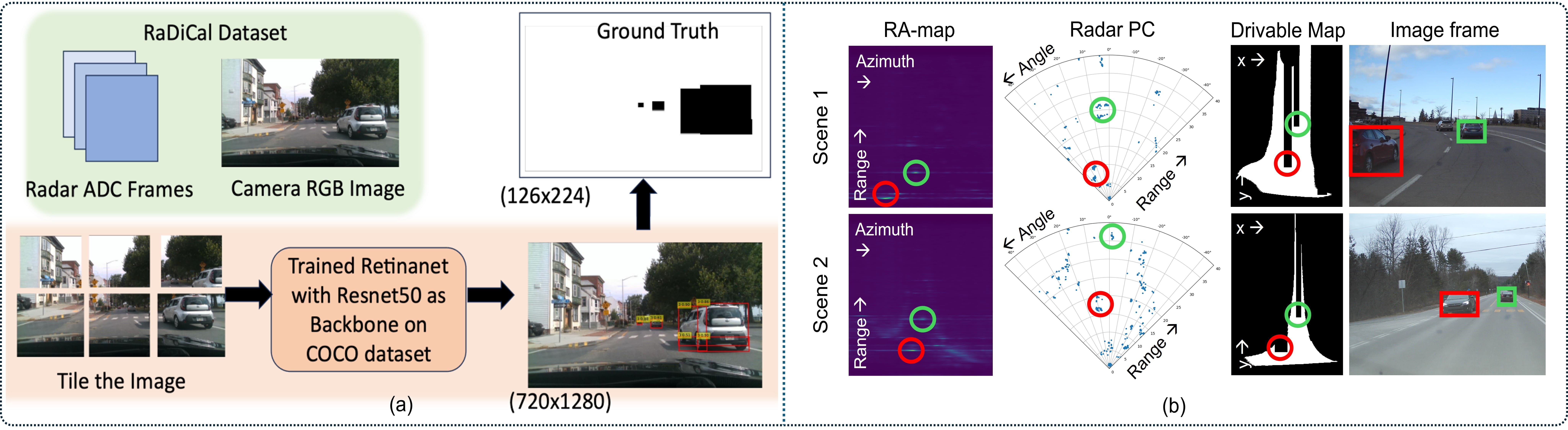}
    \caption{(a) \textbf{RaDICaL\cite{lim2021radical} Dataset:} Label generation from RGB frames (figure taken with permission from \cite{sharma2024chirpnet}). (b) \textbf{RADIal\cite{rebut2022radial} Dataset:} FFT of raw ADC data produces range–Azimuth maps; CFAR yields radar point clouds; segmentation maps mark drivable (white) vs. non-drivable (black) areas; nearest and second-nearest vehicles are highlighted in red and green consecutively.}
    \label{fig:dataset}
\end{figure*}




\subsection{BEV Projection and Decoder}\label{sec:bev_decoder}

We use similar structured decoders for both segmentation and detection tasks. The decoder takes the chirp-channel output feature map $U \in \mathbb{R}^{C\times 2N_{R_X}}$  from Chirp-SSM and passes it through a conv1D layer for dimension expansion. Then it is passed through a pooling layer across the chirp axis to accumulate the temporal information. Finally, the pooled features are reshaped to obtain the initial BEV projection ($h\times w$).

$$F_1 = \mathrm{Conv1D}(U)\,\in\,\mathbb{R}^{C \times (h\,w)}$$
$$ \bar F_1 = \mathrm{AdpAvgPool}_{C}(F_1) \in \mathbb{R}^{h w}$$
$$ F = \mathrm{Reshape}\bigl(F_1;h,w\bigr)\,\in\,\mathbb{R}^{h\times w}$$

The 2D map is then passed through upsampling layer and conv2D blocks with activation for refining spatial details for segmentation/detection.
$$Z_U^{(1)} = \mathcal U\bigl(F\bigr),\quad
    Z^{(1)} = \mathrm{SiLU}\bigl(\mathrm{Conv2D}(U^{(1)})\bigr)$$
$$Z_U^{(2)} = \mathcal U\bigl(Z^{(1)}\bigr),\quad
    Z^{(2)} = \mathrm{SiLU}\bigl(\mathrm{Conv2D}(U^{(2)})\bigr)
$$




\subsection{Why Multi‑Scale Temporal Modeling?}
For an FMCW radar, chirp samples contain information on the round-trip delay of a radio wave, containing information on the range of objects. Sample-SSM models the range across all the channels. This is important for finding the free space in the scene. However, for detection task, modeling the chirps across the radar frame is important for identifying background radio reflections and the moving objects. Chirp-SSM block helps identify these variations to identify the objects of interest in the scene. This multi-scale SSM approach enables effective separation of static background and dynamic targets in complex radar scenes.  


\section{Experimental Setup} \label{sec:exp}

This section gives an insight into the two datasets used for our experiments (RADIal\cite{rebut2022radial} and RaDICaL\cite{lim2021radical}). The metrics used for evaluation, baseline architectures for radar-only detection and the training settings.

\subsection{Datasets} \label{sec:data}

\subsubsection{RaDICaL (Radar, Depth, IMU, Camera Dataset).} RaDICaL ~\cite{lim2021radical} comprises synchronized 77 GHz FMCW radar (4 Rx × 3 Tx), stereo RGB‑D, and IMU measurements. Frame annotations were generated from synchronized camera images. To improve robustness, full‑sized images were divided into tiles~\cite{chen2019fully} for RetinaNet \cite{lin2017focalloss} inference \& detections on each tile were merged to form a binary mask, to produce ground‑truth data (\Cref{fig:dataset}(a)). Prior works have utilized image space to bird’s-eye (BEV) view mapping to convert the detections into range-azimuth space. However, we allow model to learn these linear transformation during training. The primary advantages are : (a) elimination of manual calibration across radar configurations; and (b) seamless integration with image‑space planning/control pipelines~\cite{schwarting2018planning}; and (c) It eliminates the necessity to store the radar frame in the range-azimuth plane, thereby saving storage and computational resources.

\subsubsection{RADIal (HD Radar Multi-Task Dataset).} RADIal\cite{rebut2022radial} comprises roughly two hours of synchronized driving data—front-facing RGB video, a 16-beam LiDAR and a 77 GHz imaging radar (12 Tx × 16 Rx = 192 virtual channels)—collected across 91 sequences (city streets, highways and rural roads). Of its 25 000 radar frames, 8252 are semi-automatically annotated (\Cref{fig:dataset}(b)) with per-vehicle centroids projected into the radar’s polar (range R, azimuth A) and Cartesian (X,Y) coordinates, plus a drivable-area (freespace) mask obtained by projecting the LiDAR point cloud onto the ground. For our experiment, we use a 0.8-0.2 train-validation split for our experiments.

\subsection{Implementation Details and Baselines}

We summarize the implementation details on the two datasets in \Cref{tab:dataset-hyperparams}.
\begin{table}[ht]
\footnotesize
\centering
\begin{threeparttable}
\caption{Training settings for the RADIal and RaDICaL datasets}
\label{tab:dataset-hyperparams}
\begin{tabular}{lcc}
\toprule
\textbf{Parameter}        & \textbf{RADIal} & \textbf{RaDICaL} \\
\midrule
$(C, S, N_{R_X})$         &  (256, 512, 16) &  (64, 192, 8)  \\
Training epochs           &  200            &   300           \\
Batch size                &  8              &   8               \\
Optimizer                 &  Adam           &   Adam              \\
Learning rate             &  $1\times10^{-4}$ &  $1\times10^{-4}$   \\
L2 regularization         &  $5\times10^{-6}$ & $5\times10^{-6}$   \\
Loss         &  Jaccard, Focal+Smooth L1   &  Pixel-wise BCE                \\
\bottomrule
\end{tabular}
\begin{tablenotes}
\footnotesize
\item \textbf{Note:} $C$- no. chirps/radar frame, $S$  no. samples/chirp, $N_{R_X}$ = no. receiver antennas. \textbf{BCE} - Binary cross entropy.
\end{tablenotes}
\end{threeparttable}
\end{table}

We used CNN-based (\cite{rebut2022radial,yang2018pixor,nowruzi2020deep,zhang2023adcnet,ronneberger2015unet}), Attention-based (\cite{giroux2023tfftradnet,vaswani2017attention}) \& Chirp-based recurrent (\cite{sharma2024chirpnet}) state‑of‑the‑art ML models in our experiments to establish a performance and runtime baseline for evaluation.




\subsection{Evaluation Metrics}
Following the \textbf{RADIal} benchmarking, for segmentation, we report mean intersection-over-union (\textbf{mIoU}) score between the predicted drivable-area mask and the annotated mask. In our ablation studies, we also report \textbf{dice coefficient}, mean average precision (\textbf{mAP}), mean average recall (\textbf{mAR}) and pixel-wise \textbf{accuracy} scores for testing model improvements. In the detection task, we report \textbf{mAP}, \textbf{mAR} and \textbf{F1} score.\cite{rebut2022radial}.

\begin{figure*}[t]
    \centering    \includegraphics[width=\textwidth, height = 5.5 cm]{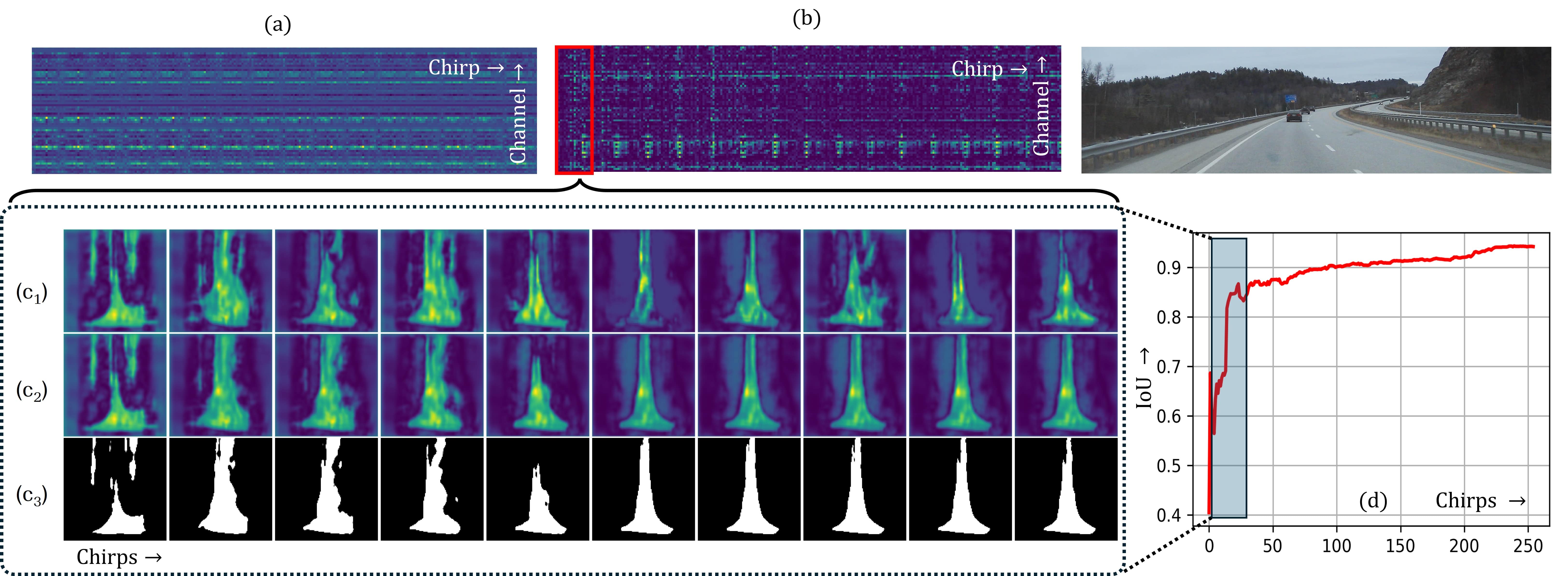}
    \caption{Feature evolution through model: (a) pooled sample-ssm output, (b) chirp-ssm output features. (c) chirp-wise feature evolution: ($\mathrm{c_1}$) independent spatial information per chirp states, ($\mathrm{c_2}$) accumulated over time, chirp state saturates, and ($\mathrm{c_3}$) a projection is formed within first 25 chirps of random initialization. (d) IoU increases rapidly at first and reaches saturation.}
    \label{fig:features}
\end{figure*}

\begin{figure*}[t]
    \centering
    \includegraphics[width = \linewidth, height = 5.2 cm]{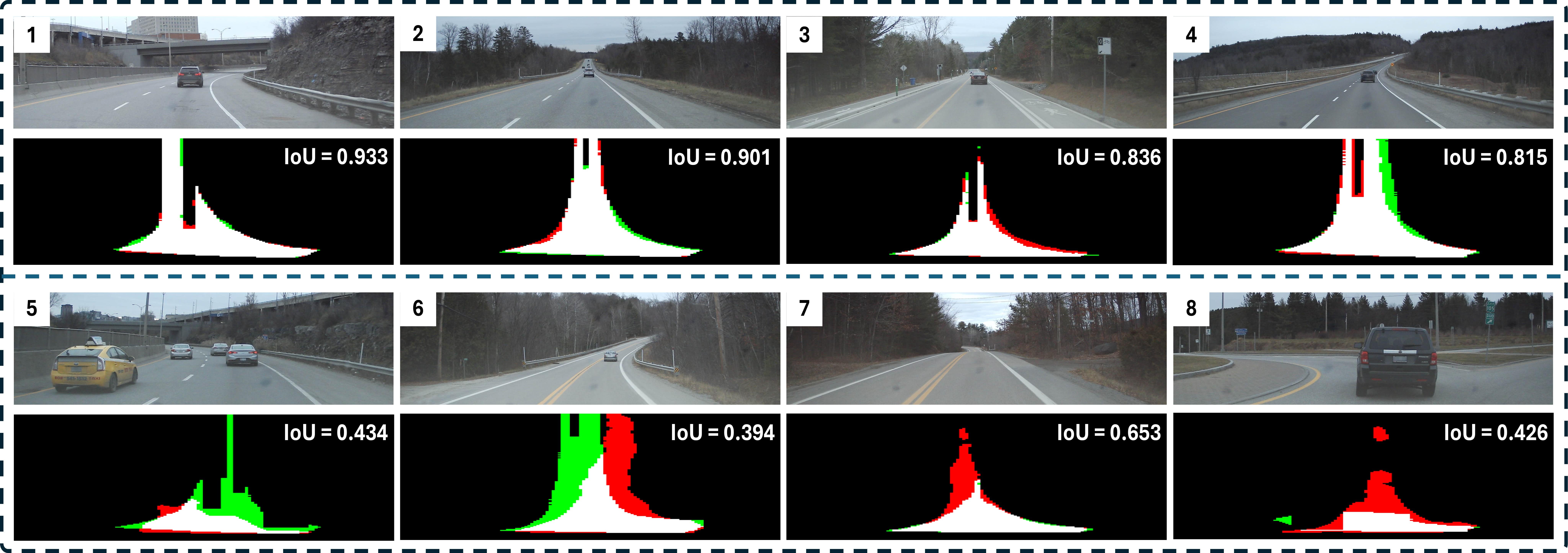}
    \caption{Ground truth vs. predicted drivable-space maps: white = true positives, green = false negatives (drivable, flagged as blocked), red = false positives (blocked, flagged as drivable). Row 1: high-IoU examples. Row 2: challenging cases (dense traffic, elevation changes).}
    \label{fig:qualitative}    
\end{figure*}

For the \textbf{RaDICaL} dataset, \textbf{Chamfer distance} was used to quantify the average bidirectional nearest‐neighbor Euclidean distance between predicted and ground‑truth BEV masks \cite{yao2023radarcamreview,fan2016point}, with lower values indicating tighter spatial alignment and thus better segmentation fidelity.  
Dice coefficient, was also reported to compare volumetric overlap between masks, with higher values indicating superior segmentation accuracy.  

Runtime performance and efficiency is assessed by multiply–accumulate operations (MACs), parameter count (M) and latency (ms) for segmentation + detection in an NVIDIA RTX 4060 mobile GPU.


\section{Results}
\subsection{Qualitative Results}

We analyze SSMRadNet’s features across stages to see how sequential radar data becomes spatial maps.
Chirpwise features are pooled into vectors, generating map \Cref{fig:features} (a). After passing through the chirp-ssm, the feature-map (b) shows distinct feature peaks and troughs.
After projecting this channel-chirp feature to spatial map followed by upsampling, \Cref{fig:features} (c) occupancy maps are formed. In this step, we compare the features extracted from each individual chirp to those obtained after aggregating chirp states across the entire frame. From ($c_2$), the spatial representation saturates early in the frame, and additional chirps contribute less to the final representation. This suggests that the radar frames can be downsized in along the chirp axis without significant loss in performance. We show some additional experiments on retaining internal state across frames in \Cref{sec:ablation}.

We compare ground truth and predicted freespace maps to reveal both successes and failure modes (Fig. \ref{fig:qualitative}). Samples 1–4 demonstrate accurate detection, with only slight drops in IoU when ground elevation fluctuates.  The bottom row illustrates more challenging cases, where errors often stem from labeling inconsistencies; since drivable-space masks were auto-generated by image/LiDAR models and not calibrated to the radar sensor (refer to \cite{rebut2022radial})—causing mismatches across elevations and obstacle distances. In sample 5, dense traffic causes the model to underestimate the free space (in green). In sample-6, 7, the ground truth and prediction maps show two different road curvatures, suggesting that the labeling might have been over constrained. 
Finally, in sample 8, the model fails to find the freespace when the view is obstructed by another vehicle at an exit, which can be attributed to edge-case samples in the dataset.
\begin{table*}[t]
\centering
\footnotesize
\resizebox{\textwidth}{!}{%
  \begin{threeparttable}
  \caption{Ablation study on \textbf{RADIal} segmentation task.}
  \label{tab:ablation}
  \begin{tabular}{lccccccccccccc}
    \toprule
    \textbf{Model} &
    \textbf{C-SSM} &
    \textbf{S-SSM} &
    \textbf{Proj} &
    \textbf{D-dim} &
    \textbf{Aggr} &
    \textbf{Expn} &
    \textbf{mIoU} &
    \textbf{Dice} &
    \textbf{mAP} &
    \textbf{mAR} &
    \textbf{Acc} &
    \textbf{Comp.} &
    \textbf{Param} \\
     &  &  &  &  &  & &
       \textbf{\%} &
       \textbf{\%} &
       \textbf{\%} &
       \textbf{\%} &
       \textbf{\%} &
       \textbf{GMACs} &
       \textbf{(M)} \\
    \midrule
    \rowcolors{1}{radStripe}{white}
    SSM based chirp model    & \checkmark & --         & \checkmark & 16 & --      & --         & 65.59 & 77.82 & 81.67 & 77.99 & 95.22 & 15.50 & 45.85 \\
    SSMRadNet-Proj           & \checkmark & \checkmark & \checkmark & 16 & Last    & --         & 67.99 & 79.88 & 82.37 & 79.99 & 95.57 & 0.73  & 13.40 \\
    SSMRadNet-NoProj-16      & \checkmark & \checkmark & --         & 16 & Last    & --         & 68.27 & 80.25 & 82.44 & 80.49 & 95.70 & 0.94  & 0.119 \\
    SSMRadNet-NoProj-32      & \checkmark & \checkmark & --         & 32 & Last    & --         & 70.59 & 81.85 & 83.62 & 82.23 & 96.07 & 1.21  & 0.126 \\
    SSMRadNet-FeatConv       & \checkmark & \checkmark & --         & 32 & Conv1D  & --         & 74.44 & 84.58 & 85.84 & 85.00 & 96.61 & 1.22  & 0.126 \\
    SSMRadNet-FeatPool       & \checkmark & \checkmark & --         & 32 & AvgPool & --         & 74.75 & 84.86 & 87.15 & 84.26 & 96.72 & 1.20  & 0.123 \\
    \rowcolor{radStripe}
    \textbf{\textcolor{blue}{SSMRadNet (Ours)}} &
      \checkmark & \checkmark & -- & 32 & AvgPool & \checkmark &
      \textbf{78.60} & \textbf{86.77} & \textbf{88.02} &
      \textbf{86.81} & \textbf{97.10} & \textbf{1.67} & \textbf{0.314} \\
    \bottomrule
  \end{tabular}
  \begin{tablenotes}
    \footnotesize
    \item \textbf{C-SSM} - Chirpwise SSM, \textbf{S-SSM} - Samplewise SSM; \textbf{Proj} - Input projection; \textbf{D-dim} - Internal vector size of Mamba-SSM D-state; \textbf{Aggr} - Feature aggregation strategy of samplewise-SSM. \textbf{\textit{Last}} - takes last state output, \textbf{\textit{Conv1D}} - 1D conv over samplewise features, \textbf{\textit{AvgPool}} - adaptive average pooling; \textbf{\textit{Expn.}} - S-SSM output channels linearly expanded to $2N_{R_X}$ before C-SSM.
  \end{tablenotes}
  \end{threeparttable}
}
\end{table*}

\begin{figure*}[htbp]
    \centering
    \includegraphics[width = 0.8\textwidth, height = 3.2 cm] 
    {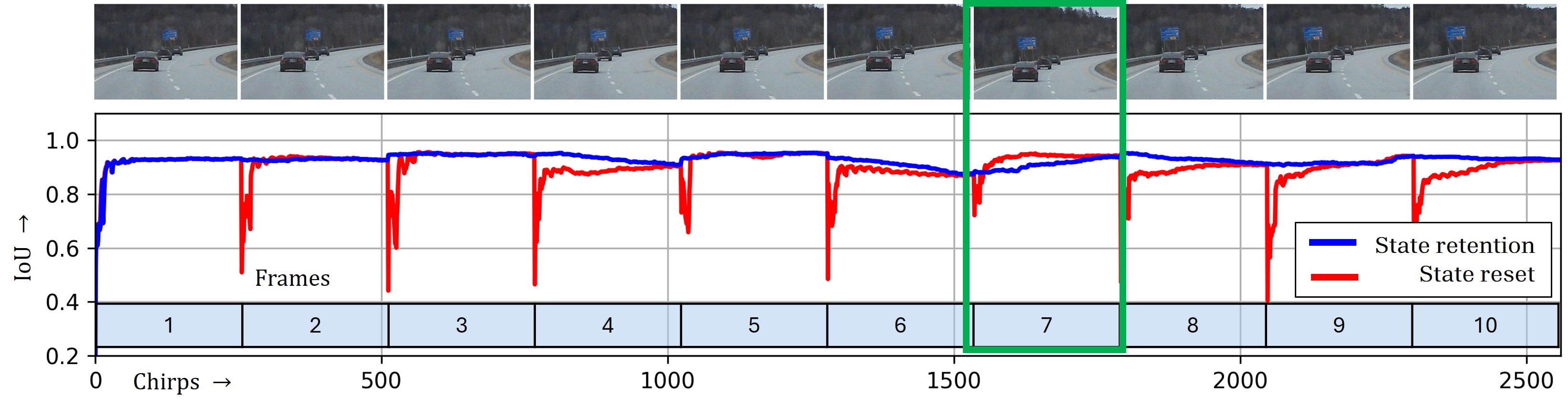}
    \caption{Decoding with state retention over frames (blue) vs state resetting per frame (red). State retention stabilizes segmentation over smooth sequences, while during physical perturbations, frame resetting performs better (green).}
    \label{fig:chirp_inference}
\end{figure*}

\subsection{Ablation Studies}\label{sec:ablation}
We begin our experiments with chirp-wise sequence modeling with SSM to reduce computation without sacrificing accuracy. In the ChirpNet\cite{sharma2024chirpnet} model family, flattened sample–channel linear projection scales poorly with large $S\times N_{R_x}$. So we introduced sample‐wise SSM that processes each sample sequentially and uses its final state as the chirp token. This change alone boosts mIoU by 2.4\% and dramatically cuts GMACs (row 2 of \Cref{tab:ablation}). Now the question comes, how to reduce the parameter count as well. We observed that projecting the entire time–channel feature map bloated parameters and destabilized training.
For this, we applied conv1D based channel expansion from encoder output, followed by a temporal pooling to convert the chirp-channel features to a 2D featuremap. This yields a slight higher segmentation gain, far fewer parameters, and more stable training.    

To test whether the SSM block D-state (output projection) has any effect in learning performance, we increased D-dim to 32 instead of 16. This gives us another 2.32\% increase in segmentation performance, with slightly increased computation. Now, we ask whether choosing only the final state output from the Chirp-SSM was a good design choice, as it restricts the model to rely more on later samples. So we experimented with feature aggregation for generating chirp tokens. Performing a feature aggregation over the sample-dimension using Conv1D (row-5)or AvgPool (row-6) after the sample-ssm  gives a huge performance increase, bringing SSMRadNet mIoU to 0.75. A last experiment on slow-time channel expansion (projecting $N_{R_X}$ to $2N_{R_X}$) before the chirp-ssm layer offered another improvement, bringing our model mIoU to 0.79, achieving SoTA performance.

\Cref{fig:chirp_inference} shows what happens if internal states are carried over to the next frame instead of resetting. In the particular scene of 10 frames, frame-wise state reset (red) yields mIoU=0.9253/frame, while state retention (blue) gives a higher score of mIoU=0.9258/frame and makes the state updates smoother. Observing each frame, we see that state retention generates early results in steady scenes, whereas a sudden change in scene (perturbations in frame 7) causes state retention to perform slightly worse than frame reset method. These observations suggest the possibility of inter-frame inferencing for speed and adaptive chirp reduction for additional computation savings.
\begin{table*}[t]
\centering
\footnotesize
\resizebox{\textwidth}{!}{%
\begin{threeparttable}
\caption{Overall segmentation and detection performance on  \textbf{RADIal \cite{rebut2022radial}}.}
\label{tab:results}
\begin{tabular}{l l c c c c c c c c c }
\toprule
\textbf{Class} 
  & \textbf{Model} 
  & \multicolumn{1}{c}{\textbf{Segmentation}} 
  & \multicolumn{5}{c}{\textbf{Detection}} 
  & \multicolumn{3}{c}{\textbf{Computational Metrics*}}\\
 &  & \textbf{mIoU} 
  & \textbf{F1} 
  & \textbf{mAP} 
  & \textbf{mAR} 
  & \textbf{RE (m)} 
  & \textbf{AE ($^{\circ}$)} 
  & \textbf{GMACs} 
  & \textbf{Params (M)} 
  & \textbf{Latency (ms)}\\
\cmidrule(lr){3-3} \cmidrule(lr){4-8} \cmidrule(lr){9-11}
\midrule

\rowcolors{1}{radStripe}{white}
\multirow{11}{*}{Convolution} 
  & Pixor (PC) \cite{yang2018pixor}                & ---  & ---  & 0.96 & 0.32 & 0.17  & 0.25  & ---   & ---   & ---     \\
  & Pixor (RA) \cite{yang2018pixor}                & ---  & ---  & 0.96 & 0.82 & 0.10  & 0.20  & ---   & ---   & ---     \\
  & PolarNet \cite{nowruzi2020deep}                & 0.61 & ---  & ---  & ---  & ---   & ---    & ---   & ---   & ---     \\
  & Conv3D + FFT-RadNet \cite{wu2024sparseradnet}  & 0.75 & 0.47 & 0.58 & 0.39 & 0.19  & 0.33   & ---   & ---   & ---     \\
  & FFT-RadNet \cite{rebut2022radial}              & 0.74 & ---  & 0.97 & 0.82 & 0.11  & 0.17   & 146.58 & 3.79  & 53.59   \\
  & FFT-RadUNet\tnote{a}                           & 0.75 & 0.80 & 0.83 & 0.77 & 0.16  & 0.09   & 134.40 & 18.48 & 44.92     \\
  & ADCNet \cite{zhang2023adcnet}                  & 0.79 & 0.89 & 0.93 & 0.86 & 0.13  & 0.11   & ---   & 2.50  & 18.13**  \\
  & ADC\,UNet \cite{zhang2023adcnet}               & 0.77 & 0.85 & 0.88 & 0.82 & 0.18  & 0.11   & ---   & 17.50 & 8.18**   \\
  & ADC\,UNet (NPT) \cite{zhang2023adcnet}         & 0.73 & 0.80 & 0.83 & 0.77 & 0.19  & 0.10   & ---   & ---   & ---     \\
  & FourierNet-FFT-RadUNet\tnote{b}                & 0.78 & 0.86 & 0.84 & 0.87 & 0.16  & 0.11   & 134.41 & 19.13 & 48.73     \\
  & FourierNet-FFT-RadNet\tnote{c}                 & 0.79 & 0.88 & 0.87 & 0.89 & 0.14  & 0.12   & 146.59 & 4.45  & 57.44     \\
\midrule

\rowcolors{1}{radStripe}{white}
\multirow{2}{*}{Attention} 
  & Self Attention-based chirp model\tnote{d}      & 0.65 & ---  & ---  & ---  & ---   & ---    & 33.00  & 50.95  & 20.37    \\
  & TFFT-RADNet \cite{giroux2023tfftradnet}        & 0.79 & 0.87 & 0.88 & 0.87 & 0.16  & 0.13   & 99.38   & 10.29  & 52.90   \\
\midrule

\rowcolors{1}{radStripe}{white}
\multirow{1}{*}{Recurrent} 
  & ChirpNet (GRU) \cite{sharma2024chirpnet}       & 0.64 & ---  & ---  & ---  & ---   & ---    & 24.70  & 57.72  & 27.33   \\
\midrule

\rowcolors{1}{radStripe}{white}
SSM & SSM-based chirp model\tnote{e}       & 0.66 & ---  & ---  & ---  & ---   & ---    & 15.50  & 45.85  & 8.32   \\
\rowcolor{radHeaderLight}
   & \textbf{\textcolor{blue}{SSMRadNet (Ours)}}    
                                         & \textbf{0.79} & 0.77 & 0.83 & 0.71 & 0.14  & 0.15
                                         & \textbf{1.67} & \textbf{0.31} & 14.20 \\
\bottomrule

\end{tabular}
\begin{tablenotes}
  \footnotesize
  \item RE = range error; AE = azimuth error; RA = range azimuth; PC = pointcloud.
  \item [a, b, c, d, e] [a] FFT-RadNet\cite{rebut2022radial} + UNet\cite{ronneberger2015unet}; FourierNet\cite{zhao2023cubelearn} FFT fed to [b] FFT-RadUNet, [c] FFT-RadNet; ChirpNet\cite{sharma2024chirpnet} with [d] transformer \cite{vaswani2017attention}, [e] SSM \cite{chen2024mamba}. 
  \item *Runtime characterization for \textbf{segmentation + detection} on an NVIDIA RTX 4060 mobile GPU. **Reported by \cite{zhang2023adcnet} on an NVIDIA RTX 3090 GPU.
\end{tablenotes}
\end{threeparttable}
}
\end{table*}

\begin{table}[htbp]
  \centering
  \scriptsize
  \setlength{\tabcolsep}{2.1pt}
  \begin{threeparttable}
    \caption{Comparison with prior works on 
     \textbf{RADICal\cite{lim2021radical}}.}
    \label{tab:radical_comparison}
    \begin{tabular}{@{} l c c c c @{}}
      \toprule
      \textbf{Model} &
      \textbf{GMACs} &
      \textbf{Params (M)} &
      \textbf{Dice Coefficient (↑)} &
      \textbf{Chamfer (↓)} \\
      \midrule
      \rowcolors{2}{radStripe}{white}
      ChirpNet \cite{sharma2024chirpnet}       & 1.480  & 3.780  & 0.986 & 0.097 \\
      ChirpNetLite \cite{sharma2024chirpnet}   & 0.320  & 3.761  & 0.989 & 0.095 \\
      ChirpNet SSM\tnote{a}                    & 0.340  & 3.761  & 0.990 & 0.088 \\
      ChirpNet-SelfAttn\tnote{b}               & 0.350  & 3.761  & 0.991 & 0.091 \\
      T-FFT-RadNet \cite{giroux2023tfftradnet} & 15.990 & 9.000  & 0.995 & 0.108 \\
      FFT-RadNet \cite{rebut2022radial}        & 41.740 & 4.250  & 0.996 & 0.076 \\
      UNet \cite{ronneberger2015unet}          & 15.140 & 17.270 & 0.996 & 0.078 \\
      \midrule
      \rowcolor{radHeaderLight}
      \textbf{\textcolor{blue}{SSMRadNet (Ours)}} & \textbf{0.108} &
                                                     \textbf{0.566} &
                                                     \textbf{0.996} &
                                                     0.086 \\
      \bottomrule
    \end{tabular}
    \begin{tablenotes}[flushleft]
      \footnotesize
      \item[a,b] [a] ChirpNet\cite{sharma2024chirpnet} with SSM; [b] with self-attention \cite{vaswani2017attention}.
      \item * Dice coefficient higher is better; Chamfer Distance lower is better.
    \end{tablenotes}
  \end{threeparttable}
\end{table}

\subsection{Results on RaDICaL}
SSMRadNet achieves a dice score of \textbf{0.996} on the \textsc{RaDICaL} segmentation task (\Cref{tab:radical_comparison}), matching the top‑performing FFT-RadNet \cite{rebut2022radial}, U‑Net\cite{ronneberger2015unet} models, outperforming T-FFT-RadNet\cite{giroux2023tfftradnet}, while requiring \(\mathbf{148\times}\) less compute and \(\mathbf{16\times}\)–\(\mathbf{30\times}\) fewer parameters. 
Despite the drastic reduction in GMACs and model complexity, SSMRadNet maintains state‑of‑the‑art segmentation accuracy, achieving an average Chamfer Distance of \textbf{0.086}, comparable with FFT-RadNet and U‑Net, while surpassing transformer‑based T‑FFT‑RadNet baselines and other ChirpNet variants.

\subsection{Results on RADIal}
SSMRadNet attains \textbf{0.79} mIoU on the segmentation task, showing competitive performance with state-of-the-art models like T-FFTRADNet\cite{giroux2023tfftradnet} with $60\times$ less compute (1.67 GMACs) and $33\times$ fewer parameters (0.33 M) than \cite{giroux2023tfftradnet} (\Cref{tab:results}), while beating benchmark models like FFTRadNet\cite{rebut2022radial} and ChirpNet\cite{sharma2024chirpnet}. Detailed results on the segemntation task is provided in \Cref{tab:results}, listing all the SoTA models grouped according to architecture type discussed in \Cref{tab:comparison}. 
In the detection task, SSMRadNet has yet to achieve SoTA performance as shown in \Cref{tab:results}, achieving F1 score of 0.77, mAP score of 0.83 and mAR score of 0.71. Simplified decoder structure used in our experiments is one of the key factors behind the performance loss in detection, and is a possible direction for future work.



\section{Conclusion \& Future Work}\label{sec:conc}
We presented SSMRadNet, a compact neural architecture that processes raw radar ADC signals via sample-wise and chirp-wise state-space models to produce BEV occupancy maps. Across two diverse datasets, SSMRadNet matches or exceeds prior state-of-the-art radar-only methods in segmentation benchmarks while using less than one million parameters and only a few GFLOPs per frame. Despite a minor decrease in detection performance, it remains competitive with other SOTA radar-only approaches. By processing sample‑wise raw ADC sequences with hierarchical state space modeling architecture, SSMRadNet opens a new direction for efficient multi‑task radar perception. 

Future work could focus on training with adverse weather data to stress test model robustness. 
Moreover, SSMRadNet naturally lends itself to multi-modal fusion: additional camera or LiDAR embeddings could be incorporated into the SSMs or fused in the BEV decoder, combining radar’s all-weather sensing with rich visual context. Finally, the linear-scaling compute of SSMs makes our approach suitable for next-generation radars with higher chirp count and larger antenna arrays. We hope this work inspires further exploration of lightweight, radar-specific deep learning architectures for autonomous systems.
\section{Acknowledgement}
\label{sec:ack}
This material is based upon work supported in part by SRC JUMP~2.0 (CogniSense, \#2023-JU-3133) and in part by DARPA through the OPTIMA program. Any opinions, or recommendations expressed in this material are those of the author(s) and do not reflect the views of SRC or DARPA.
 
\newpage
{
    \small
    \bibliographystyle{ieeenat_fullname}
    \bibliography{main}
}

\end{document}